%% file: main.tex
\documentclass[conference]{IEEEtran}
\usepackage{amsmath}
\usepackage{algorithm}
\usepackage{cite}
\usepackage{stfloats}
\usepackage{multicol}
\usepackage{multirow}
\usepackage{tabularx}
\usepackage{longtable}
\usepackage{booktabs}
\usepackage{array}
\usepackage{url}
\usepackage{nomencl}
\usepackage{cite}
\usepackage{amssymb,amsfonts}
\usepackage{graphicx}
\usepackage{textcomp}
\usepackage{xcolor}
\usepackage{soul}

\usepackage{algpseudocode}
\usepackage{amsmath}
\usepackage{flushend}
\usepackage{enumitem}
\usepackage{comment}
\usepackage[hidelinks]{hyperref}
\usepackage{soul}
\begin{document}
% \title{Optimization of Dynamic Line Rating and Energy Storage System placements}

\title{Two-Stage Optimization for Dynamic Line Rating and Energy Storage Deployment \vspace{-0.25em}}

\author{\IEEEauthorblockN{Abanish Tiwari,~\emph{Student Member, IEEE}, Phurba Sherpa,~\emph{Student Member, IEEE}, Chandan Chaudhary, \\ ~\emph{Student Member, IEEE}, Mohammed Benidris,~\emph{Senior Member, IEEE}, and Joydeep Mitra,~\emph{Fellow, IEEE}}
\IEEEauthorblockA{Electrical and Computer Engineering, Michigan State University, East Lansing, MI 48824, USA}
Emails: tiwariab@msu.edu, sherpaph@msu.edu, chaud152@msu.edu, benidris@msu.edu, and mitraj@msu.edu \vspace{-1.25em}}

\maketitle

% This is the abstract Section
\begin{abstract}
\input{0_Abstract}
\end{abstract}

\IEEEpeerreviewmaketitle
% This is the introduction section
\section{Introduction}
\input{1_Introduction}
\vspace{-0.5em}
\section{Methodology}
\label{sec:methodology}
\input{2_Methodology}

% This has the results and discussions
\section{Test System Implementation}
\label{sec:test_system_implementation}
\input{3_ResultsandDiscussion}
%\vspace{-1em}
% The conclusion goes here
\section{Conclusion}
\label{sec:conclusion}
\input{4_Conclusion}

\section*{Acknowledgment}
The research is supported in part by U.S. National Science Foundation (NSF) under Grant No. 2404872 and in part by computational resource provided by ICER at Michigan State University.

\bibliography{references}
\bibliographystyle{ieeetr}

\end{document}

%% file: 0_Abstract.tex
The increasing penetration of distributed energy resources (DER) and weather-driven variability has intensified congestion and reliability stress in transmission networks. Strategies that enhance the utilization of existing infrastructure, such as static line ratings (SLR) and energy storage systems (ESS), have therefore become necessary. SLRs rely on conservative ambient assumptions and often understate thermal limits, whereas dynamic line ratings (DLR) adjust capacity according to weather conditions and unlock additional transfer capability. Energy storage systems provide temporal flexibility, but their transmission-level effectiveness depends on proper siting and sizing. This paper proposes a two-stage optimization method for joint placement of DLR installations and utility-scale energy storage. In the first stage, a mixed-integer linear program selects DLR corridors and ESS buses by minimizing operating cost, DER curtailment, and load-shedding penalties subject to DC power flow and investment constraints. In the second stage, the model determines ESS energy capacity and operating schedules under ambient-driven line ratings. Ambient weather data is used to generate DLR profiles, and sequential Monte Carlo simulation is applied to assess system adequacy. The proposed method, when deployed on the modified IEEE RTS 24-bus system, shows that coordinated DLR and ESS planning improves transmission capability, mitigates congestion, and strengthens system adequacy under weather variability.

%Monte Carlo simulation is used to evaluate system reliability, weather data ---> historical

%% file: 1_Introduction.tex
The electric power system is witnessing rapid structural change driven by the proliferation of distributed energy resources (DERs), large-scale electrification, and the emergence of highly concentrated, high-demand loads such as artificial intelligence (AI) data centers. These developments place unprecedented operational stress on transmission networks, where capacity constraints, congestion, and thermal bottlenecks increasingly limit the deliverability of low-cost energy and the reliable integration of DERs. In the US, recent assessments indicate that more than 1,350 GW of prospective clean generation and storage capacity, which is roughly equivalent to the clean power needed for an 80\% clean electricity share by 2030, now awaits transmission interconnection, yet completion rates remain below 25\% amid rising wait times and upgrade costs \cite{doeInterconnection2024, doetransmission2024}. As building new transmission systems remains slow due to permitting, environmental, and siting barriers, enhancing the utilization of existing grid infrastructure has become essential to realize the economic, operational, and reliability benefits of the evolving resource mix.

Static line ratings (SLR) are based on worst-case ambient assumptions and therefore, they substantially under-represent the actual thermal capability of transmission lines throughout most operating hours \cite{ieee738_2023}. As a result, grid operators are forced to curtail DER output, redispatch higher-cost generation, and maintain larger reserve margins even when real-time weather conditions would safely permit higher power transfers. Dynamic line ratings (DLR) address these limitations by adjusting thermal limits based on actual or forecasted ambient temperature, wind speed, wind direction, and solar irradiance. A wide body of work demonstrates that DLR can unlock substantial latent transmission capacity, reduce congestion costs, and enhance deliverability of low-carbon resources~\cite{glaum2023leveraging, lee2022impacts, abdelkader2026td, stuehm2002dynamic}. The benefits of DLR, however, hinges on strategic deployment, as localized environmental conditions and network interdependencies determine which corridors yield the greatest operational benefit.

Energy storage systems (ESS) offer complementary flexibility by absorbing surplus DER output, which relieves loading on constrained interfaces, and support the system during high-demand or adverse weather conditions \cite{Sanchez2024Ess, arteaga2021energy}. ESS also mitigates DER variability and reduces power imbalances by charging during periods of excess generation and discharging when supply is scarce. Optimal siting and sizing of ESS is therefore a critical planning task, but existing research often focuses either on cost minimization, curtailment reduction, or distribution-level applications, without fully capturing composite system reliability impacts \cite{Sanchez2024Ess, kumar2022role, zhao2015optimal}. Likewise, the value of DLR, depends on where it is deployed. System-wide installation is impractical due to sensor costs, communication requirements, and cybersecurity considerations. Several studies have proposed methods for selecting DLR lines based on congestion patterns, wind integration levels, thermal overloading or reliability impacts ~\cite{abdelkader2026td, glaum2023leveraging, Teng2018}.

 By exploiting favorable environment conditions, DLR expands transmission capability while ESS strengthens operational flexibility during periods of stress, and their coordinated use enables the grid to accommodate more DERs, enhance reliability, and defer major network reinforcements \cite{glaum2023leveraging, abdelkader2026td}. Despite these complementary benefits, most existing studies evaluate DLR and ESS independently, without accounting for their operational coupling or their combined impact on transmission-level reliability under weather-driven uncertainty.

This paper addresses this gap by proposing a coordinated, two-stage optimization method that simultaneously determines the optimal placement of DLR installations, ESS locations, and ESS capacity. In stage~1, the model identifies the transmission corridors where DLR provides the greatest operational benefit, along with the most effective bus for ESS placement, based on operating cost, curtailment penalties, and load-shedding minimization. In stage~2, the ESS capacity and charge--discharge schedules are optimized to enhance flexibility and reliability under weather-dependent line ratings. Weather-driven DLR profiles are generated using Sequential Monte-Carlo simulation (SMCS), and system reliability is evaluated across baseline and coordinated deployment scenarios.

The remainder of the paper is organized as follows. Section \ref{sec:methodology} presents the proposed two-stage optimization framework in detail including the reliability indices considered for evaluation. Section \ref{sec:test_system_implementation} implements and validates the proposed framework in IEEE RTS-24 bus test system and analyses the results. Finally, Section~\ref{sec:conclusion} provides concluding remarks and outlines directions for future work.  

%% file: 2_Methodology.tex
\begin{figure}[!htbp]
    \centering 
    \vspace{-1cm}
    \includegraphics[width=0.5\textwidth]{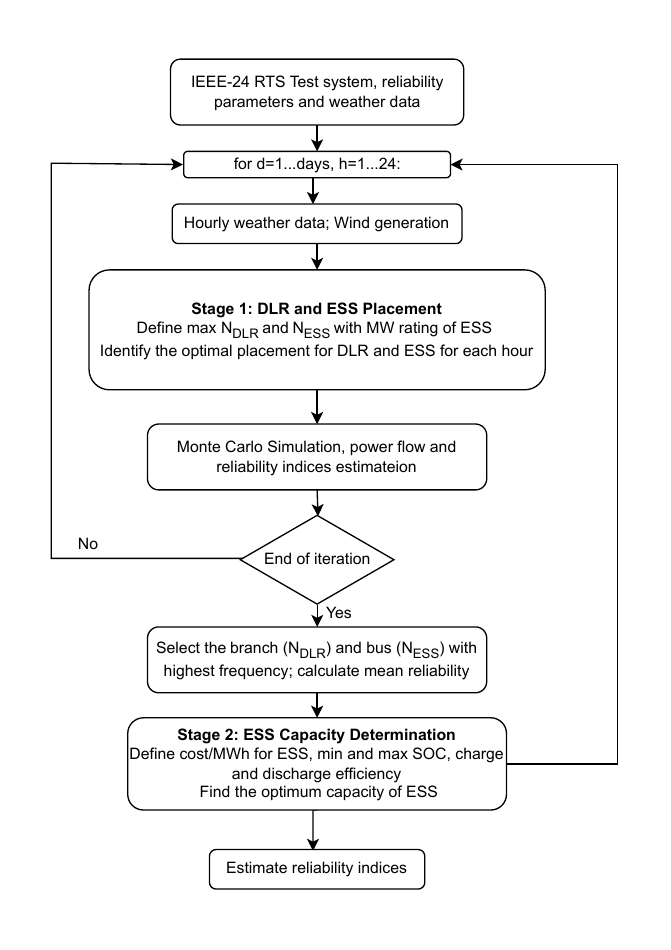}
    \vspace{-1.2 cm}
    \caption{Flowchart of the proposed methodology}
    \vspace{-1.5em}
    \label{fig:methodology}
\end{figure}

This section presents the proposed approach to determine the DLR and optimal placement, sizing, and operation of DLR installations and ESS under weather-dependent system conditions. 
Fig. \ref{fig:methodology} illustrates a detailed overview of the methodology adapted in this study.

% The framework developed presents the optimized location for the placement of DLR and ESS, minimizing the operational costs and power curtailment and shedding limited by the budget constraint. The Mixed Integer Linear Programming (MILP) is used as the optimization tool with open-source CBC solver for decisions of DLR and ESS placement. The DC optimal power flow (OPF) is used to evaluate the system performance. This section explains the methodology followed in detail.

\subsection{Dynamic Line Rating Calculation}
Dynamic line ratings are calculated using the IEEE Std. 738-2023 guidelines \cite{ieee738_2023} for conductor thermal ratings. The steady-state conductor rating is determined using the heat balance equation:
\begin{equation}
q_c(T_c, T_a, V_m, \phi) + q_r(T_c, T_a) = q_s(S_I) + I_{DLR}^2 R(T_c)
\label{heatEq}
\end{equation}
where $T_c$ and $T_a$ represent the conductor and ambient temperatures, $V_m$ and $\phi$ denote the wind speed and its direction, $S_I$ is the solar radiation, $I_{max}$ and $R$ are the conductor current and resistance, $q_c$ and $q_r$ are the convective and radiative cooling, and $q_s$ is the solar heating. By setting the conductor temperature $T_c$ to its thermal limit, the rated current $I_{DLR}$ can be calculated for given ambient conditions. This allows the determination of dynamic thermal limits $S_{ij,t}^{DLR}$ for each transmission line during periods of elevated loading to benefit from DLR. 
Historical data: $T_c$, $T_a$, $V_m$ and $\phi$, considered the same throughout the system, obtained from \cite{NSRDB} are used evaluate the new DLR ratings from \eqref{heatEq} for each branch, which are $\alpha_{ij,t}$ times the previous rating.
% Historical hourly data obtained from \cite{nsrdb_website} were considered throughout the overall system.

% \subsection{Load and Generation sources}
% The hourly load data was obtained from the IEEE RTS 24-bus \cite{grigg1999}. The load is considered to be increased by 1.8 times the existing load. Similarly, the maximum capacity of the generation sources increased by 1.5 times as the wind sources are integrated into IEEE-RTS system \cite{wind paper here}. The generation from the wind rources was considered from
\subsection{Wind Generation}
Wind generation is modeled following the probabilistic approach in \cite{wind_power_model}. Wind speed is obtained from \cite{NSRDB}, and turbine output is defined through its characteristic power curve using the cut-in ($V_{ci}$), rated ($V_r$), and cut-out ($V_{co}$) wind speeds. For a given wind speed realization $V_m$, the turbine output is zero below $V_{ci}$ and above $V_{co}$, increases with wind speed between $V_{ci}$ and $V_r$, and remains at rated power $P_r$ between $V_r$ and $V_{co}$.

Hourly wind speeds are mapped to power output states through this curve to construct a discrete probability distribution of wind generation. This probabilistic representation enables integration of wind power into reliability evaluation and adequacy assessment as described in \cite{wind_power_model}.

% The electrical power output of a wind turbine is modeled as a piecewise function of the wind velocity $V_m$, following the turbine power-curve characteristics as presented in \cite{wind_power_model} and is expressed as
% \begin{equation}
% P(V_m) =
% \begin{cases}
% 0, & 0 \le V_m \le V_{ci}, \\
% (A + B V_m + C V_m^{2}) P_r, & V_{ci} \le V_m \le V_r, \\
% P_r, & V_r \le V_m \le V_{co}, \\
% 0, & V_m \ge V_{co},
% \end{cases}
% \label{eq:power_curve}
% \end{equation}
% where $P_r$ is the rated power of the wind turbine.$V_{ci}$, $V_r$ and $V_{co}$ represent the cut-in, rated, and cut-out speed of the wind turbine. The coefficients A, B \& C are defined in \cite{wind_power_model}.
% \begin{equation*} 
% \scalebox{1}{$ 
% \begin{aligned} A &= \displaystyle \frac{1}{(V_r - V_{ci})^{2}} \left[\, V_{ci}(V_{ci}+V_r) - 4V_{ci}V_r - \frac{(V_r+V_{ci})^{2}}{2V_r} \,\right],
% \end{aligned} $}  
% \end{equation*}
% \hl{can we take 2 out of equation in C = ?}
% \begin{align}
%     B &= \frac{1}{(V_r - V_{ci})^{2}} \left[\, 4(V_{ci}+V_r) - \frac{(V_r+V_{ci})^{3}}{2V_r^{2}} - 3(V_r+V_{ci}) \,\right]\notag, \\[4pt] 
% C &= \frac{1}{(V_r - V_{ci})^{2}} \left[\, 2 - 4\,\frac{(V_{ci}+V_r)^{2}}{2V_r} \label{eq:wind_coeffs},\right] 
% \end{align}
% and,
% $V_w$ is the wind velocity and $P_r$ is the rated power of the wind turbine. $V_{ci}$, $V_r$ and $V_{co}$ represent the cut-in, rated, and cut-out speed of the wind turbine.

% \subsection{Decision Variables}
\subsection{Optimization Problem Formulation}
This section formulates the optimization problem to determine the optimal placement of DLR and ESS. The problem is formulated as a mixed-integer linear programming (MILP) that minimizes the operational cost, power curtailment and load shedding subjected to operational and budgetary constraints. A DC power flow is employed within each scenario to compute dispatch, line flows, and node-level imbalances under weather-dependent DLR limits and ESS operation. The power flow outcomes are then used to evaluate the reliability indices. 
% The following sections describe the two-stage optimization method in detail:

% --okay % Only stage 1 is completed, 
\subsubsection{Decision Variables} 
The investment variables include the binary variable $x_{ij} \in \{0,1\}$ that indicates DLR installation on line $(i,j)$, and the binary variable $y_k \in \{0,1\}$ that indicates the ESS installation at bus $k$ for Stage 1. In Stage 2, for each selected ESS site, the decision variable  $E_k^{ESS}\ge 0$ determines its  energy capacity (MWh). 
%$P_k^{ESS}\ge 0$ and power rating (MW) and

\subsubsection{Operational Variables} 
The operational variables include the scenario-dependent system behavior, i.e, ESS charging and discharging power $(P^{ch}_{k,t}, P^{dis}_{k,t})$, state of charge (SOC) $E^{SOC}_{k,t}$, DER curtailment $P^{curt}_{g,t}$, bus voltage angles $\theta_{k,t}$, and power flows $(P_{ij,t})$. \(u_t\) is a binary mode-selection variable that enforces mutual exclusivity by allowing the ESS to either charge or discharge at any given time step, but not both.

This determines the optimal placement of DLR installations and ESS units subjected to operational, thermal, and investment constraints.

%Stage~2 refines the design by determining the ESS energy capacity and enforcing realistic switching behavior between charging and discharging modes. The complete mathematical formulation for both stages is provided below:
% The upper level decision variables include the 
% binary variable $x_{ij} \in \{0,1\}$ which indicates DLR installation on line $(i,j)$, and the binary variable $y_k \in \{0,1\}$ that indicates the ESS installation at bus $k$. the power rating $P_k^{ESS} \geq 0$ (MW), and energy capacity $E_k^{ESS} \geq 0$ (MWh). Operational variables include ESS charging power $P_{k,t,s}^{ch} \geq 0$ (MW), discharging power $P_{k,t,s}^{dis} \geq 0$ (MW), and state of charge $E_{k,t,s}^{SOC} \geq 0$ (MWh). Renewable curtailment is captured by $P_{g,t,s}^{curt} \geq 0$ (MW). Power system operational variables include bus voltage angles $\theta_{k,t,s} \in \mathbb{R}$ (rad), line active power flows $P_{ij,t,s} \in \mathbb{R}$ (MW), and apparent power flows $S_{ij,t,s} \geq 0$ (MVA).

\subsection{Stage 1: Problem Formulation}
The most beneficial transmission lines for DLR deployment and the optimal buses for ESS installation are determined by minimizing the total operating cost, DER curtailment, and load shedding across all scenarios. The optimization is formulated as MILP in which siting decisions are represented by binary investment variables, while system operations are governed by scenario-dependent dispatch and power flow variables.

\subsubsection{Objective Function}
% \vspace{-0.3em}
\begin{equation}
\begin{aligned}
\min \!&
\sum_{t\in\mathcal{T}} \! \left[
\sum_{g\in\mathcal{G}} C_g P_{g,t}
\!+\! C^{\mathrm{curt}}\!\sum_i p^{\mathrm{curt}}_{i,t}
\!+\! C^{\mathrm{shed}}\!\sum_i p^{\mathrm{shed}}_{i,t}
\right]\!
\end{aligned}
\label{eq:s1_obj_switch1}
\end{equation}
which aggregates the generator cost $C_g$, associated penalties: curtailment cost $C^{curt}$ and unserved demand cost $C^{shed}$.

\subsubsection{Constraints}
The constraints associated with the objective function are discussed below:
\begin{enumerate}[label = \alph*.]
    \item DC power flow:
    DC power flow is implemented through a linearized relationship between active power and voltage angle differences, with a designated slack bus to guarantee solvability.
    \begin{equation}
    P_{ij,t}=\frac{\theta_{i,t}-\theta_{j,t}}{X_{ij}},\qquad \theta_{\mathrm{ref},t}=0
    \label{eq:s1_dc}
    \end{equation}

     \item Nodal power balance:
    Nodal power balance ensures that the generation, ESS operation, DERs output, and load collectively match the net power injections and withdrawals across connected lines.
    \begin{equation}
    \begin{aligned}
    \sum_{g\in\mathcal{G}(i)}P_{g,t}
    + P^{ESS}_{i,t}
    + (P^R_{i,t}-p^{\mathrm{curt}}_{i,t})- P^D_{i,t}\\
    = \sum_{j:(i,j)\in\mathcal{L}} P_{ij,t}
    + p^{\mathrm{shed}}_{i,t,s}, \quad \forall i,t.
    % \sum_{g\in\mathcal{G}(i)}P_{g,t}
    % + (P^{\mathrm{dis}}_{i,t}-P^{\mathrm{ch}}_{i,t})
    % + (P^R_{i,t}-p^{\mathrm{curt}}_{i,t})\\
    % - P^D_{i,t}
    % = \sum_{j:(i,j)\in\mathcal{L}} P_{ij,t}
    % + p^{\mathrm{shed}}_{i,t,s}, \quad \forall i,t.
    \end{aligned}
    \label{eq:s1_balance}
    \end{equation}

    \item Generator and curtailment bounds:
    Generation must remain within the physical operating limits, and DERs curtailment cannot exceed the resources available at each bus at that time.
    \begin{equation}
    \begin{split}
     P_g^{\min} &\le P_{g,t}\le P_g^{\max},\\
     0 &\le p^{\mathrm{curt}}_{i,t}\le P^R_{i,t}.
    \end{split}
    \label{eq:s1_gen}
    \end{equation}

    \item Load shedding bounds:
    The load shedding bounds is given by:
    \begin{equation}
    0 \le p^{\mathrm{shed}}_{i,t} \le P^D_{i,t},
    \label{eq:s1_shed}
    \end{equation}
    which restricts unserved demand to be non-negative and not exceed the total load at each bus at given time.
    
    \item Thermal limits with DLR:
    Line limits adjust dynamically based on installed DLR equipment and scenario-specific ambient conditions captured through the factor $\alpha_{ij,t,s}$.
    \begin{equation}
    \begin{aligned}
     -P^{\mathrm{max}}_{ij,t}\le P_{ij,t}\le P^{\mathrm{max}}_{ij,t},\\
     P^{\mathrm{max}}_{ij,t}= 
     {P}^{rated}_{ij}[(1-x_{ij})+x_{ij}\alpha_{ij,t}\big].
    \end{aligned}
    \label{eq:s1_dlr}
    \end{equation}

    \item Investment limits: Deployment of DLR installations and ESS units is restricted by maximum allowable quantities $N_{\mathrm{DLR}}^{\max}$ and  $N_{\mathrm{ESS}}^{\max}$ respectively to represent practical investment limitations.
    \begin{equation}
    \sum_{(i,j)\in\mathcal{L}} x_{ij}\le N_{\mathrm{DLR}}^{\max},\qquad
    \sum_{i\in\mathcal{B}^{\mathrm{ESS}}} y_i\le N_{\mathrm{ESS}}^{\max},
    \label{eq:s1_budgets}
    \end{equation}
    
    % \begin{equation}
    % 0\le P^{\max}_i \le y_i\,\overline{P}^{\mathrm{site}},\quad \forall i\in\mathcal{B}^{\mathrm{ESS}}.
    % \label{eq:s1_pmax}
    % \end{equation}
    \end{enumerate}

    \subsection{Stage 2: Problem Formulation}
    Stage~2 determines the optimal ESS energy capacity for the optimal locations from Stage 1 and incorporates additional constraints associated with ESS dynamics to assess their reliability contribution under weather-dependent and DLR installed conditions.
    \subsubsection{Objective Function}
    The objective function is given by
    % Obejctive function for SoC
    \vspace{-1em}
    \begin{equation}
    \begin{aligned}
    \min \; & 
    \sum_{t\in\mathcal{T}} \Bigg[
        \sum_{b\in\mathcal{B}} E^{ESS} \, C_{\text{cap}}
        + \sum_{g\in\mathcal{G}} C_g P_{g,t} \\
    & \qquad
        + C^{\mathrm{curt}} \sum_i p^{\mathrm{curt}}_{i,t}
        + C^{\mathrm{shed}} \sum_i p^{\mathrm{shed}}_{i,t}
    \Bigg]
    \end{aligned}
    \label{eq:s1_obj_switch2}
    \end{equation}
    %\vspace{-1em}
    this aggregates the generator cost and  per MWh cost of ESS $C_{cap}$ with penalties.
    \subsubsection{Constraints}
    In addition to the constraints in \eqref{eq:s1_dc}-\eqref{eq:s1_budgets}, the following additional constraints are considered in Stage~2:
    \begin{enumerate}[label = \alph*.]
    \item  SOC Evolution:
    The SOC of the ESS evolves in time according to the net effect of charging and discharging actions, adjusted by their respective efficiencies $\eta^{ch}$ and $\eta^{dis}$, carried forward from the previous time step.
    \begin{equation}
    E^{SOC}_{k,t}
    =
    E^{SOC}_{k,t-1}
    + \eta^{ch} P^{ch}_{k,t}\,\Delta t
    - \frac{1}{\eta^{dis}} P^{dis}_{k,t}\,\Delta t,
    \label{eq:soc_evolution}
    \end{equation}
    
    \item No simultaneous charge and discharge:
    Any energy storage cannot charge and discharge simultaneously. 
    \begin{equation}
    \begin{aligned}
    %& P^{ch}_{t} \cdot P^{dis}_{t} = 0, \\ the equation below also insures no charging and discharging at a time
    & P^{ch}_{t} \le M\,u_{t}, P^{dis}_{t} \le M(1-u_{t}), \quad 
      u_{t} \in \{0,1\}.
    \end{aligned}
    \end{equation}
    where \(M\) is a sufficiently large constant. In this case, larger than the maximum charge/discharge rating of the ESS, \(M \ge \max P^{ESS}_k, \forall k\).

    \item{ESS power limits}: Charge and discharge power for each ESS unit remains within its rated power capability:
    \begin{equation}
    0 \le P^{ch}_{k,t} \le P^{ESS}_{k}, \quad 0 \le P^{dis}_{k,t} \le P^{ESS}_{k}, \quad \forall k,t.
    \end{equation}

    \item{SOC capacity bounds:}
    The SOC of each ESS unit is restricted within minimum and maximum fractions of its installed energy capacity:
    \begin{equation}
    \mathrm{soc}_{\min}\,E^{\max}_k 
    \le 
    E^{\mathrm{SOC}}_{k,t}
    \le 
    \mathrm{soc}_{\max}\,E^{\max}_k,
    \quad \forall k,t,
    \label{eq:soc_bounds}
    \end{equation}
    where \(\mathrm{soc}_{\min}\) and \(\mathrm{soc}_{\max}\) denote the minimum and maximum SOC fractions.
\end{enumerate}

\vspace{-0.75em}
\subsection{Reliability Indices}System adequacy is assessed using standard reliability indices computed from the outcomes of SMCS. These indices quantify the likelihood, frequency, and magnitude of supply shortfalls as a result of insufficient available capacity under weather-dependent operating conditions.

\subsubsection{Loss of Load Probability (LOLP)}
The LOLP quantifies the probability that the system is unable to meet load demand in a given scenario and is expressed as:
% \begin{equation}
% LOLP = \frac{1}{N_s} \sum_{s=1}^{N_s} \mathbb{I}[\text{load shedding occurs in scenario } s]
% \end{equation}
\begin{equation}
\mathrm{LOLP}
= \frac{1}{N_s}
  \sum_{s=1}^{N_s}
  \mathbb{I}\!\left[
    \sum_{k \in \mathcal{B}}
    \sum_{t \in \mathcal{T}}
    P^{\mathrm{shed}}_{k,t,s} > 0
  \right],
\end{equation}
where $\mathbb{I}[\cdot]$ is the indicator function that returns 1 when load shedding occurs and 0 otherwise.

\subsubsection{Loss of Load Expectation (LOLE)}
The LOLE measures the expected number of time intervals during which load shedding occurs, expressed in hours per year:
% \begin{equation}
% LOLE = \sum_{s \in \mathcal{S}} \sum_{t \in \mathcal{T}} \pi_s \mathbb{I}[P_{k,t,s}^{shed} > 0 \text{ for any } k \in \mathcal{B}] \times \frac{8760}{T}
% \end{equation}
\begin{equation}
\mathrm{LOLE}
=
\sum_{s \in \mathcal{S}}
\sum_{t \in \mathcal{T}}
\pi_s\,
\mathbb{I}\!\left[
\sum_{k \in \mathcal{B}}
P^{\mathrm{shed}}_{k,t,s} > 0
\right]
\times \frac{8760}{T},
\end{equation}
where $\pi_s$ is the probability of scenario $s$, and the factor $8760/T$ converts simulated time steps into expected hours per year.

\subsubsection{Expected Unserved Energy (EUE)}
The EUE quantifies the total expected energy not served to customers over a year and is computed as:
\vspace{-0.75em}
\begin{equation}
\mathrm{EUE}
=
\sum_{s \in \mathcal{S}}
\sum_{k \in \mathcal{B}}
\sum_{t \in \mathcal{T}}
\pi_s\, P^{\mathrm{shed}}_{k,t,s}\, \Delta t
\times \frac{8760}{T\,\Delta t},
\end{equation}
where $\Delta t$ is the duration of each time interval and the scaling factor converts simulated unserved energy into annual MWh.

The load shedding term, $P^{\mathrm{shed}}_{k,t,s}$, is obtained hourly from DC OPF executed in each scenario.
% , and line flows are calculated using voltage-angle differences and line reactances correspondingly from \eqref{eq:s1_dc}.

%% file: 3_ResultsandDiscussion.tex
The proposed method is validated in a modified IEEE-RTS test system from \cite{Sanchez2024Ess}, with high penetration of wind resources to introduce weather-driven variability and operational uncertainty. System parameters were adjusted to intensify the stress on generation and network assets, thereby creating conditions under which DLR and ESS coordination provides measurable benefits. Specifically, the system load was scaled by a factor of 1.8 and the generation capacity was increased by a factor of 1.5 to reflect high-demand, high-renewable penetration conditions. All MILP formulations are implemented in the \texttt{PuLP} optimization toolkit with the \texttt{COIN-OR Branch and Cut (CBC)} mixed-integer solver \cite{pulp} in the High Performance Computing Cluster of Michigan State University. System adequacy is evaluated through 500{,}000 SMCS iterations. The MILP is solved for the binary variables $x_{ij}$ and $y_k$ to determine the {$n_{ij} \times 1$} (branch) and {$k \times 1$} dimensional array to determine the DLR and ESS positions for the minimum objective function. The decision for the DLR and ESS is made based on the highest number of occurrences of the particular bus or branch.
% \vspace{-1em}
\begin{figure}[!htbp]
    \centering 
    \includegraphics[width=0.5\textwidth]{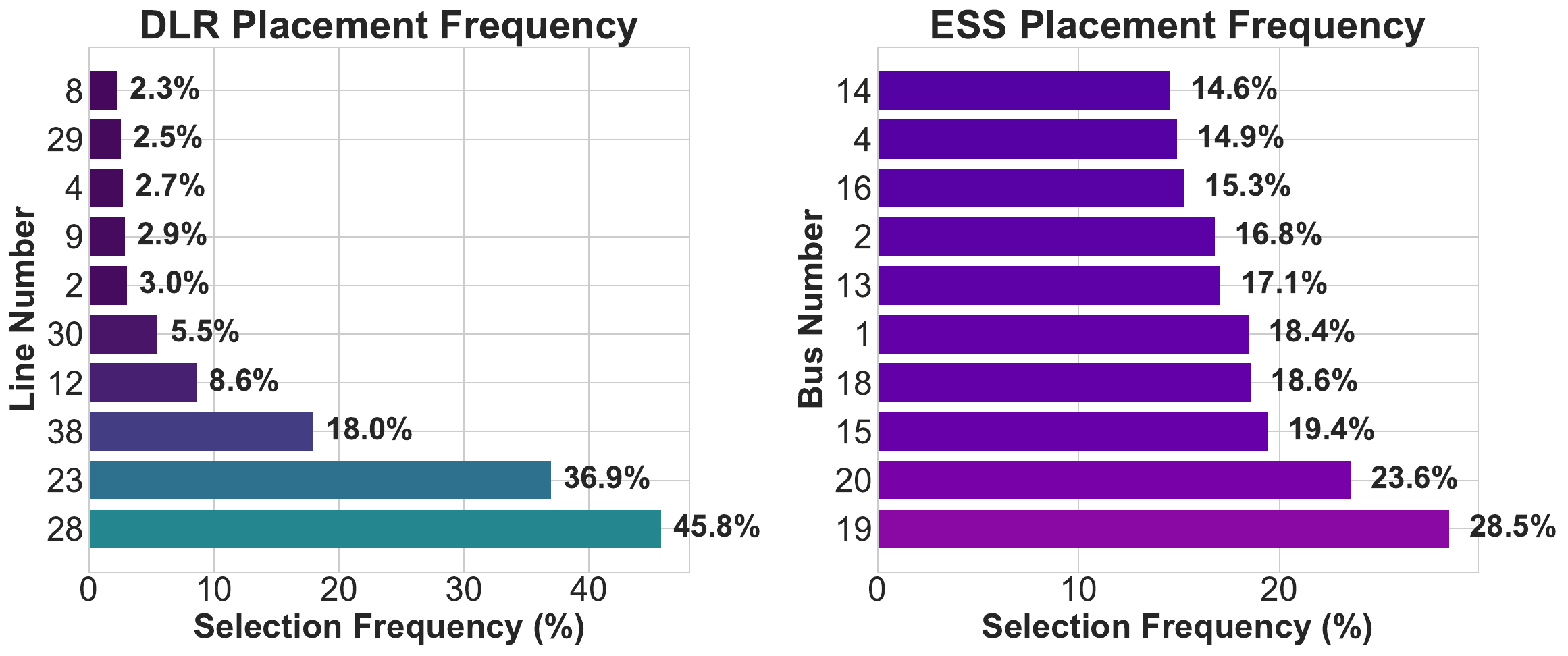} 
    \caption{Top 10 decisions for DLR \& ESS placement}
    \vspace{-0.35 cm}
    \label{fig:dlr_bess_prob}
\end{figure}

Fig.~\ref{fig:dlr_bess_prob} summarizes the most frequently selected DLR-enabled transmission branches and ESS installation buses across all simulations. Budget constraints permitted a maximum of three DLR installations and two ESS sites, while the Stage~2 optimization determined the ESS energy capacity of 534.38~MWh for the specified power rating of 150~MW. The identified bus for optimal placement of DLR and ESS is tabulated in Table~\ref{tab:Placement_results}. The resulting investment decisions highlight a clear preference for corridors experiencing both high congestion and strong sensitivity to weather-driven ampacity variations, illustrating the importance of coordinated siting under uncertainty.

\begin{table}[!htbp]
\centering
\vspace{-1.25em}
\caption{DLR and ESS placement}
\label{tab:Placement_results}
\begin{tabular}{lcc}
\hline
\textbf{Description} & \textbf{Branch} & \textbf{Bus} \\
\hline
DLR & 28, 23, 38 & - \\
% \hline
ESS & - & 19, 20 \\
\hline
\end{tabular}
% \vspace{-0.75em}
\end{table}

Table~\ref{tab:Reliability_results} reports the reliability performance of the system before and after optimization. Coordinated DLR--ESS planning yields substantial improvements across all adequacy indices. The LOLP decreases from 0.121 to 0.061 with an improvement of 49.6\%. The LOLE is nearly halved, from 44.043 h/year to 22.267 h/yr,  with an approximate improvement of 49.4\%. The EUE improved by 55\% with the reduction of the unserved energy reduced by 8,749.85 MWh/year. These results confirm that targeted DLR deployment combined with optimally sized storage markedly strengthens system adequacy under stressed, weather-dependent operating conditions.

\begin{table}[!htbp]
\centering
\vspace{-0.75em}
\caption{Impact of DLR–ESS Coordination on Reliability Indices}
\label{tab:Reliability_results}
\begin{tabular}{lrrr}
\hline
\textbf{Metric} & \textbf{Base} & \textbf{After Opt} \\
\hline
LOLP & 0.121 & 0.061 \\
LOLE (h/year) & 44.043 & 22.267 \\
EUE (MWh/year) & 15,903.424 & 7,153.574 \\
\hline
\end{tabular}
% \vspace{-0.3cm}
\end{table}

\begin{figure}[!htbp]
    \centering 
    \vspace{-0.25em}
    \includegraphics[width=0.48\textwidth]{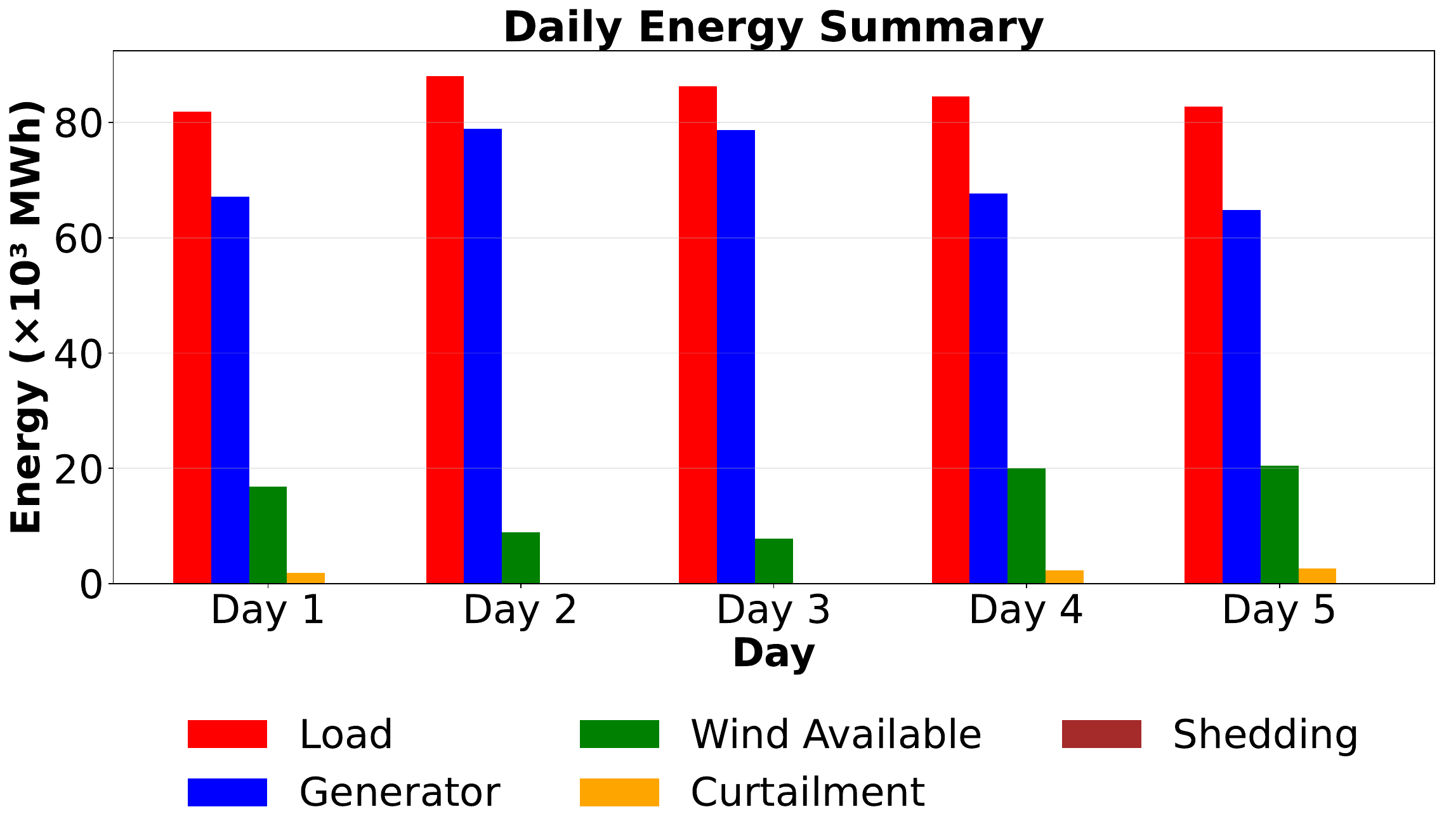}
    \vspace{-0.75em}
    \caption{Representative energy profile}
    \vspace{-0.25cm}
    \label{fig:ener_flow}
\end{figure}

\begin{figure}[!htbp]
    \centering 
    \includegraphics[width=0.5\textwidth]{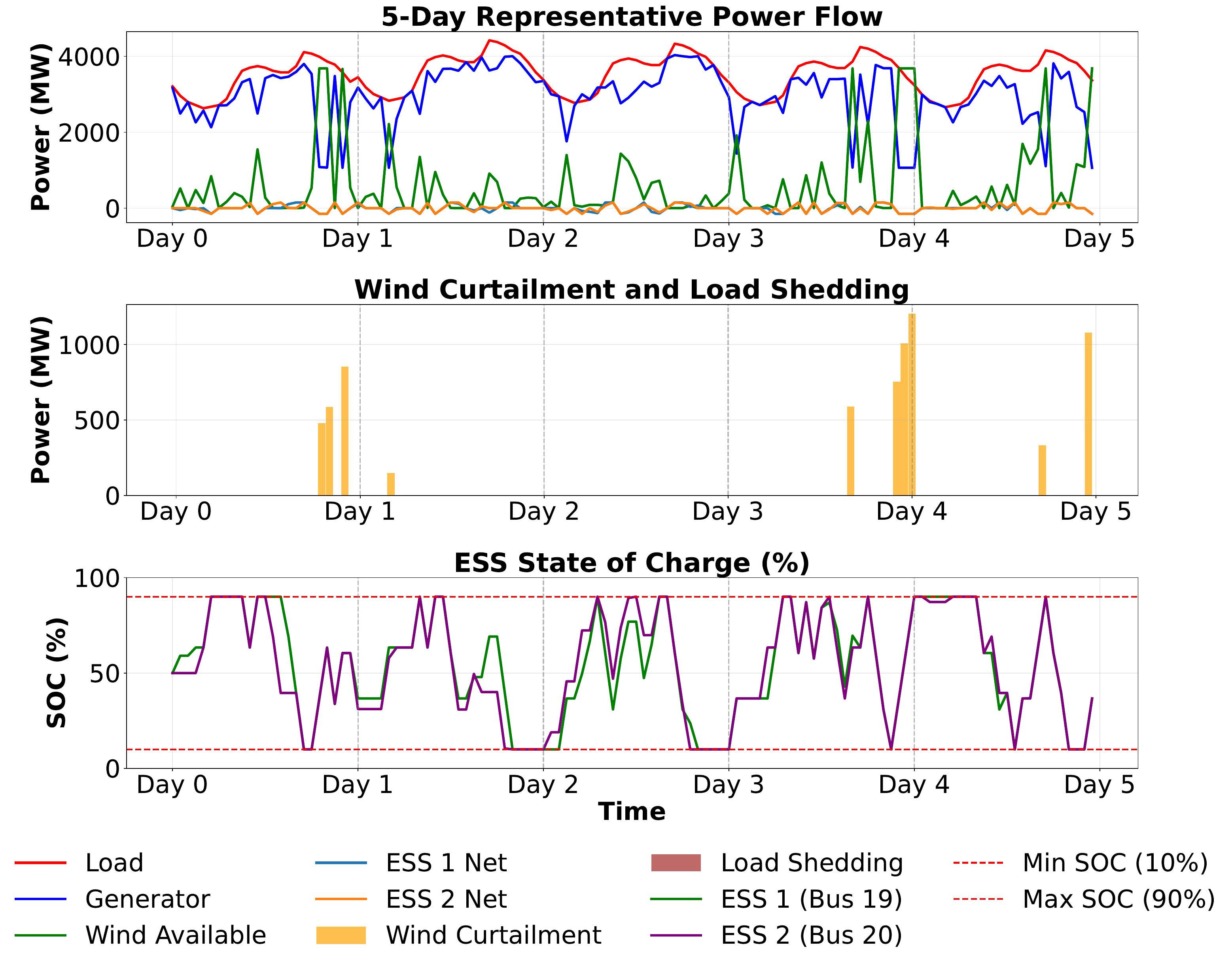}
    \vspace{-0.5cm}
    \caption{Representative power flow, curtailment and SOC profile}
    \label{fig:power_flow}
    \vspace{-0.5cm}
\end{figure}

The representative operational pattern of the ESS units appears in Figs. \ref{fig:ener_flow} and~\ref{fig:power_flow}. Wind generation exerts a strong influence on the net system load, thus both ESS units display similar charge and discharge trajectories. They absorb surplus wind during high-speed intervals and supply power during wind lulls. This behavior, supported by the optimized 150~MW/534.38~MWh sizing, enables the storage fleet to buffer renewable variability and reduce reliance on conventional generation during stressed periods.

% \vspace{-0.4cm}
%\vspace{-0.5cm}

The results indicate that coordinated planning of DLR and ESS provides improvements in system adequacy by simultaneously addressing transmission congestion and renewable-driven variability. The consistently selected transmission branches indicate branches where thermal limits are highly sensitive to weather-dependent ampacity and frequently experience congestion under stressed operating conditions. Furthermore, the bus for ESS installation corresponds to the locations where storage most effectively mitigates the supply–demand imbalances. The observed improvements in reliability indices confirm the value of the coordinated approach. As storage absorbs surplus wind generation and supplies power during wind shortfalls, it complements the congestion relief provided by DLR. This synergy between increased transmission capability and temporal energy shifting enhances the ability of the system to maintain supply–demand balance under uncertainty. Overall, the results validate that the proposed co-optimization problem solving approach improves system adequacy by explicitly capturing the interaction between network constraints and renewable variability within a unified model.

%% file: 4_Conclusion.tex
The paper has proposed a coordinated, two-stage optimization method for the optimal placement and sizing of the DLR installations and ESS under weather-dependent operating conditions. By integrating ambient-driven DLR limits, ESS temporal flexibility, and probabilistic reliability evaluation within a coordinated plan for placement and ESS sizing. The method identifies the plan that meaningfully enhances system adequacy. Simulation results on a modified IEEE RTS test system demonstrate substantial improvements across all major reliability indices. The findings confirm that coordinated DLR-ESS planning yields benefits higher than each technology can achieve independently, especially in the systems with high DER penetration and weather-driven variability.

Future research includes the incorporation of correlated weather fields and short-term forecasting to better represent spatiotemporal dependencies. Multi-period expansion planning to jointly optimize transmission upgrades, storage sizing and siting, and DER deployment can be performed. Applying the framework to larger network models and evaluating the computational acceleration techniques would further establish its scalability for real-world planning studies.  